\documentclass[aps,pra,superscriptaddress,twocolumn,amsmath,amsfonts,amssymb,floatfix]{revtex4-1}
\usepackage{enumerate}
\usepackage{mathtools}
\usepackage{amsmath}
\usepackage{graphicx}
\usepackage{epsfig}
\usepackage{sidecap}
\usepackage{hyperref}
\usepackage{color}
\usepackage{subfigure,array}
\usepackage{verbatim}

\usepackage{enumerate}
\usepackage{bm}

\begin{document}

\title{Dropleton-Soliton Crossover mediated via Trap Modulation}

\author{Argha Debnath}
\author{Ayan Khan} \thanks{ayan.khan@bennett.edu.in}
\affiliation{Department of Physics, School of Engineering and Applied Sciences, Bennett University, Greater Noida, Uttar Pradesh, India}
\author{Saurabh Basu}
\affiliation{Department of Physics, Indian Institute of Technology - Guwahati, Guwahati, Assam, India}

\begin{abstract}
We report a droplet to a soliton crossover by tuning the external confinement potential in a dilute Bose-Eienstein condensate by numerically solving the modified Gross-Pitaevskii equation. The testimony of such a crossover is presented via studying the fractional density of the condensate which smoothly migrates from being a flat-head curve at weak confinement to a bright soliton at strong confinement. Such a transition occurs across a region of the potential whose strength varies over an order of magnitude and thus should be fit to be termed as a crossover. We supplement our studies via exploring the size of the bound pairs and the ramifications of the particle density therein. Eventually, all of these aid us in arriving at a phase diagram in a space defined by the trap strength and the particle number that shows the formation of two phases consisting of droplets and solitons, along with a regime of coexistence of these two.
\end{abstract}

\maketitle

\section{Introduction}
The three states of matter, namely, solid, liquid and vapor are integral part of our daily experience. However, in certain conditions other states of matter can also emerge, such as plasmas at high temperatures, and superfluids at very low temperatures. These different phases are manifestation of distinct interactions between their constituent particles. For example, in a regular liquid, we observe the competition between van der Waals inter particle attraction and the short-range repulsion due to the Pauli exclusion principle. The transition from a liquid to a solid phase is also observed by lowering the temperature. Interestingly, quantum liquids may not exhibit similar properties. For example, helium remains a liquid close to absolute zero temperature (under atmospheric pressures) due to the significant zero-point motion of the atoms. At this low temperature quantum mechanical effects are the dominating factor, which supports the emergence of superfluidity. Nevertheless, they can still poses some characteristics similar to classical liquid such as droplet formation \cite{bottcher}. 

The experimental observation of atomic Bose-Einstein condensate (BEC) \cite{anderson,hulet,ketterle} had paved the way for the emergence of new experimental platforms \cite{stringari1999,stringari2008,bloch}. This has allowed us to study quantum phenomena and exotic states of matter with precise controllability and tunability \cite{grimm}. 
As a result, lately liquid-like state in a two component BEC \cite{cabrera1} has been reported. This unique development influences us to investigate the exotic phase more closely as the existing understanding of liquid formation has considerable influence from the theory of Van der Waals. Yet, these newly emerged droplets in dilute ultra-cold atomic gases do not explicitly follow the familiar notion of classical liquids \cite{barbut2}. These are truly quantum liquids where quantum fluctuation plays an important role \cite{barbut1,barbut3}.
These droplets are small clusters of atoms self-bound by the interplay of attractive and repulsive forces emerging from the mean-field (MF) and the beyond mean-field (BMF) interactions \cite{sala1}. The BMF contribution is nothing but the Lee-Huang-Yang's (LHY) correction \cite{lee} of the Gross-Pitaevskii (GP) equation \cite{gross,pitae}. In a two component BEC, the effective MF interaction can be significantly suppressed due to the competitions between the intra-species and inter-species interactions. This may result in a situation when the effective MF interaction strength is in the order of LHY correction and competing in nature. This leads to an ideal scenario for droplet formation. The experimental verification was based on the theoretical proposition where it was claimed that, if the square of the inter-species coupling is greater than the product of the intra-species coupling then the collapse of the two-component BEC can be averted and a dilute liquid-like droplet state emerges \cite{petrov}. 

However it must be noted that, the first experimental observation of droplet was on dipolar condensate of $^{164}Dy$ \cite{barbut1,barbut3} with subsequent theoretical developments coming to light \cite{santos}. Later the droplets were observed for a binary BEC comprised of two hyper-fine states in $^ {39} K$ \cite{cabrera1,fattori}. Later, soliton to dropleton transition was also reported \cite{cabrera2}. Quantum droplets were also observed in a heteronuclear mixture of bosonic gases \cite{fort}. In a recent self-consistent theoretical formulation LHY correction is incorporated in the GP equation via quantum fluctuation \cite{sala2}. We have also noted theoretical assertions on the collective modes in a droplet-soliton crossover \cite{sala1}, existence of vortices in droplets \cite{malomed,lin,chen,zhang}, dynamics of purely one-dimensional droplets \cite{astra}, its collective excitations \cite{tylutki} along with comprehensive reviews \cite{luo}. 
At this juncture, we also recall a numerical investigation of quantum liquids for the dipolar BEC in quasi one-dimensional (Q1D) geometry \cite{edmonds}. Apart from that, we have also seen considerable interest in analyzing the origin of the droplets in lower dimensions \cite{chen,malomed_review}. A contemporary study also notes possible connection between the droplets and modulational instability in an one dimensional system \cite{khare}.

Very recently, we have proposed an exact analytical solution for the droplet-like state in a Q1D quasi-homogeneous cigar shaped BEC where we have assumed the longitudinal confinement is sufficiently weak so that it can be neglected \cite{debnath1}. In that article we have also commented on the soliton to droplet transition. However, from an experimental perspective, it is more prudent to consider the cigar shaped harmonic confinement in Q1D as depicted in Fig.~\ref{cartoon}. Tuning the confinement potential should yield additional features that may be worth exploring \cite{dey}. So, we start with an expectation of finding some novel consequences by tuning the trapping potential. 
Our systematic analysis reveals that, the trap frequency indeed plays a significant role in understanding the droplet-soliton transition.     

In this letter, we consider a binary BEC of two hyper-fine states while both components occupy the same spatial mode \cite{cabrera1}. As a result, the two-component nonlinear Schr\"odinger equation (NLSE) is reduced to an effective one component equation.
Next, we employ a dimensional reduction reduction scheme following the prescription of Ref.\cite{atre}. This allows us to reduce the three dimensional system to a quasi one-dimension without loss of much generality. Physically, a dimensional reduction makes sense, when the transverse components of the anisotropic harmonic confinement is much stronger compared to the longitudinal trapping frequency. The Q1D system now consists of a cubic and a quartic nonlineairty, contrary to the prevailing concept of NLSE where we generally deal with only a cubic nonlinearity. The cubic term defines the effective mean-field (EMF) two-body interaction and the quartic term is the signature of beyond mean-field contribution. We have recently carried out a systematic analysis for cubic-quartic nonlinear Schr\"odinger equation (CQNLSE) \cite{debnath1} in quasi one dimension and reported exact analytical solutions. Further, we have analyzed and portrayed the existence of cnoidal functions \cite{abro} in the CQNLSE when subjected to cnoidal confinements \cite{debnath2}. Lately, we have 
investigated on the existence of periodic modes in CQNLSE and commented on the dimensional crossover from Q1D to an one dimensional system \cite{debnath3}. In this letter, our main objective is to focus on the role played by the trap potential in droplet formation and possible transition from a droplet to a solitonic state via tuning the confinement potential and studied numerically by solving the inhomogeneous CQNLSE.

In this paper, we communicate our results in the following order, in Sec.\ref{model}, we review the necessary theory which maps the two component BEC to one component extended Gross-Pitaevskii (GP) equation. Thereafter we explicate the formalism to reduce the three dimensional system to Q1D geometry. We briefly recap the scheme to derive the analytical solution and compare them with our current numerical code in Sec.\ref{sol}. In Sec.\ref{drop}, first we investigate the effect of the longitudinal trapping potential for fixed number of particles at different values of the BMF interaction strength. Later we extend the analysis toward the role of particle number and draw phase diagram. This allows us to comment on the nature of droplet to soliton transition. We draw our conclusion in Sec.\ref{con}.

\section{Theoretical Model}\label{model}
\begin{figure}
\includegraphics[scale=0.25]{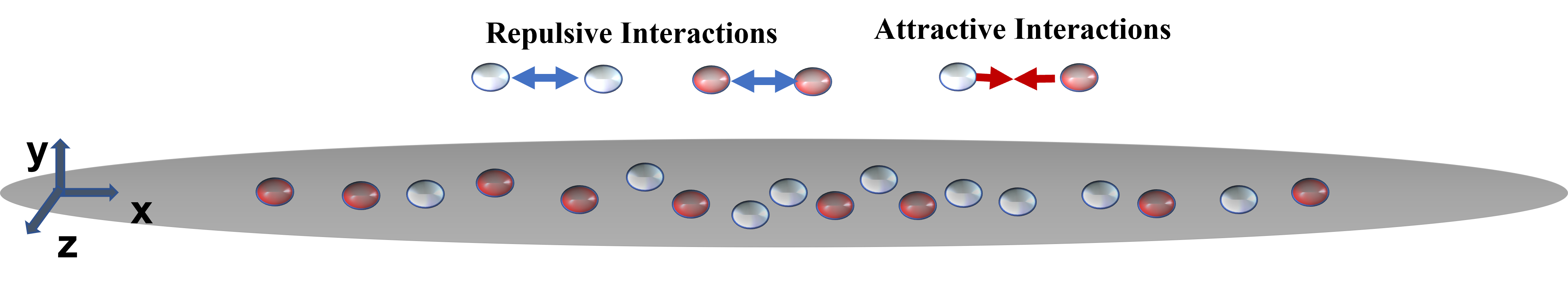}
\caption{(Color online) Pictorial illustration of binary BEC in a Q1D confinement. The two different hyper-fine states are described by the bluish and reddish spheres. The intra-species and inter-species interactions are considered to be repulsive and attractive in nature.}\label{cartoon}
\end{figure}
Here, we consider a mixture of two hyper-fine states of $^{39}K$ in a Q1D geometry following the prescription of Ref.\cite{cabrera1}. 
A schematic description of the physical system can be visualized in Fig.~\ref{cartoon} where the bluish and the reddish spheres depict the two different hyper-fine states of atoms
distributed in an effectively cigar shaped or Q1D trap. The intra-species interactions ($a_{11}$ and $a_{22}$) are characterized as repulsive and the inter-species interaction ($a_{12}$ or $a_{21}$) is considered as attractive. The experimentally observed droplets are described as small clusters of atoms self-bound through the competition between the attractive and the repulsive forces. 

It is important to mention at this juncture that, there exists notable research on purely one dimensional (1D) systems \cite{astra}. However, we know that condensation is not feasible in one dimension and even though the energy of a weakly interacting Bose gas can be predicted using Bogoliubov theory \cite{lieb1,popov,petrov1}. Thus Q1D geometry is more amenable from experimental perspective. This leads to a significant deviation in beyond mean field contribution with regard to the purely 1D scenario. From a mathematical perspective, the current model navigates us toward CQNLSE whereas the 1D system is described via quadratic-cubic NLSE (QCNLSE) \cite{petrov1}.

Assuming both the components, of the described binary condensate, occupy the same spatial mode, it is possible to express the system via an effective single component GP like equation. 
Thus the resulting equation of motion, with cubic and quartic nonlinearities is noted as \cite{cabrera1,debnath1},
\begin{eqnarray}
i\hbar\frac{\partial\Phi}{\partial t}&=&\left[\left(-\frac{\hbar^2}{2m}\nabla^2+V_{Ex}\right)+U_1|\Phi|^2+U_2|\Phi|^3\right]\Phi,\nonumber\\
&&\label{3dbgp}
\end{eqnarray}
where, $U_1$ and $U_2$ denotes the effective mean-field and BMF interaction strength,
$m$ being the mass of the atoms and
$V_{Ex}$ is the external potential which is expressed as a combination of the transverse and longitudinal components of the harmonic confinement. 
In cigar-shaped BEC or in Q1D geometry transverse trap frequency is typically $10$ times larger than
the longitudinal frequency. This ensure that the interaction energy of the atoms remains much lesser compared to the kinetic energy in the transverse direction.

Now, one can reduce Eq.(\ref{3dbgp}) to an effective one-dimensional equation. The dimensional reduction is performed by employing the following ansatz, 
\begin{eqnarray}
\Phi(\mathbf{r}, t) &=& \frac{1}{\sqrt{2\pi a_B}a_{\perp}}\phi\left(\frac{x}{a_{\perp}},\omega_{\perp}t\right) e^{\left(-i\omega_{\perp}t-\frac{y^2+z^2}{2a_{\perp}^2}\right)},\nonumber\\\label{ansatz1}
\end{eqnarray}
where, $a_B$ is the Bohr radius, $a_{\perp}=\sqrt{\frac{\hbar}{m\omega_{\perp}}}$ and $\omega_{\perp}$ is the transverse trapping frequency.

Inserting the ansatz of Eq.(\ref{ansatz1}) in Eq.(\ref{3dbgp}), we obtain the extended GP equation as noted below,
\begin{align}
i\frac{\partial\phi(x,t)}{\partial t} = &  \left[ - \frac{1}{2}\frac{\partial^2}{\partial x^2} + \frac{1}{2} K x^{2}+G_1 |\phi(x,t)|^2\right.\nonumber\\&\qquad\left.+\frac{ }{ }G_2 |\phi(x,t)|^3  \right]\phi(x,t)\label{bgp},
\end{align}
where, $G_1$ and $G_2$ are appropriately scaled mean-field and BMF interaction strength in Q1D. Additionally, the trapping frequency in longitudinal direction ($\omega_0$) is now scaled as 
$K = \omega^{2}_{0}/\omega^{2}_{\perp}$.
Further, it must be noted that $x$ and $t$ are in units of $a_{\perp}$ and $\omega_{\perp}^{-1}$. 
From here onward, we shall follow this dimensionless notations of $x$ and $t$ i.e., $x\equiv x/a_{\perp}$ and $t\equiv\omega_{\perp}t$. 

In this letter, our main focus is to explicate the role of $K$ in dropleton-soliton transition. For this purpose, we solve Eq.(\ref{bgp}) numerically for different values of $K$. The numerical scheme involved a split-step Crank-Nicolson (CN) method with imaginary time propagation \cite{muru}. However, we prefer to briefly revisit our analytical solution for better readability. 

\section{Solution}\label{sol}
To obtain an analytical solution, we start our analysis from the ansatz solution, $\phi(x,t)=\eta(x,t)\exp{\left[i\left(\Sigma(x,t)+\mu_0 t\right)\right]}$, where $\eta(x,t)$ leads to the amplitude contribution and $\Sigma(x,t)$ is the non-trivial phase, $\mu_0$ being the chemical potential \cite{debnath1}. After some algebra, we arrive at the stable analytical solution in the center of mass frame as \cite{debnath1},
\begin{eqnarray}
\eta(\zeta)&=&\frac{1+12\mu_g}{1+\sqrt{12\mu_g}\cosh{\left(\sqrt{g_1}\zeta\right)}}\,\textrm{for}\, |g_1|=3g_2. \label{con2}
\end{eqnarray}
Here, $\zeta=x-ut$ described the comoving frame with $u$ being the velocity of the condensate. In the analytical derivation we have also assumed that $g_1=-2G_1$ and $g_2=2G_2$ implying a two-body effective mean-field interaction is attractive and LHY contribution is repulsive and $\mu_g=\mu_0/g_1$. 

The relationship between the normalization, $N$ and
chemical potential, $\mu_0$ can be noted as \cite{debnath1},
\begin{eqnarray}
N&=&\frac{(1+\mu_I)^2}{\sqrt{g}(1-\mu_I)}\left[\ln\left[1-\frac{2}{\mu_I}\left(\sqrt{1-\mu_I}+1\right)\right]-2\right],\label{N}
\end{eqnarray}
where, $\mu_I=12\mu_g$ and $N$ relates to the particle number associated with the formation of localized wave. 

\begin{figure}
\includegraphics[scale=0.3]{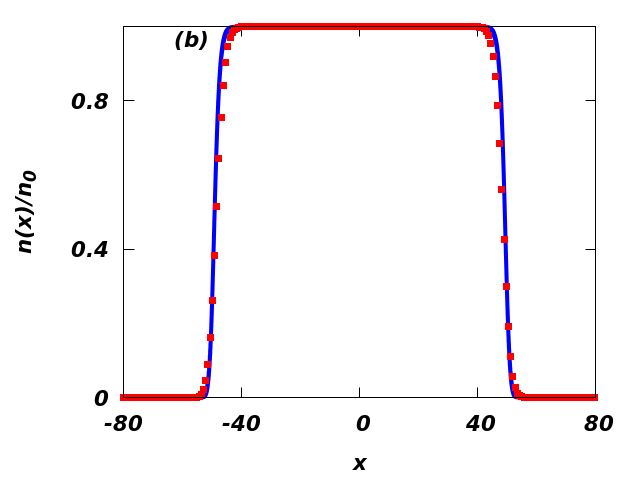}
\caption{(Color online) The figure depicts the comparison between the analytically obtained solution as noted in Eq.(\ref{con2}) and numerical solution of Eq.(\ref{bgp}) (with $K=0$). 
The analytical solution is plotted with solid blue line whereas the red squares represents numerically obtained solution.  The density is normalized by $n_0$ where $n_0$ is $n(x)|_{x=0}$.}\label{nu_an}
\end{figure}

We test our numerical scheme, by comparing with the analytical result as depicted in Fig.~\ref{nu_an}. In the figure, the solid blue line is our analytical solution from Eq.\ref{con2} and the red solid squares are the numerically obtained solution of Eq.(\ref{bgp}) for $K=0$. Fig.~\ref{nu_an} corresponds to the droplet like state. The computation appears in good agreement with the analytical result and thus the scenario encourages us to proceed towards our actual objective of solving Eq.(\ref{bgp}) with $K\neq0$. 

\section{Droplet to Soliton Transition}\label{drop}
\begin{figure}
\includegraphics[scale=0.35]{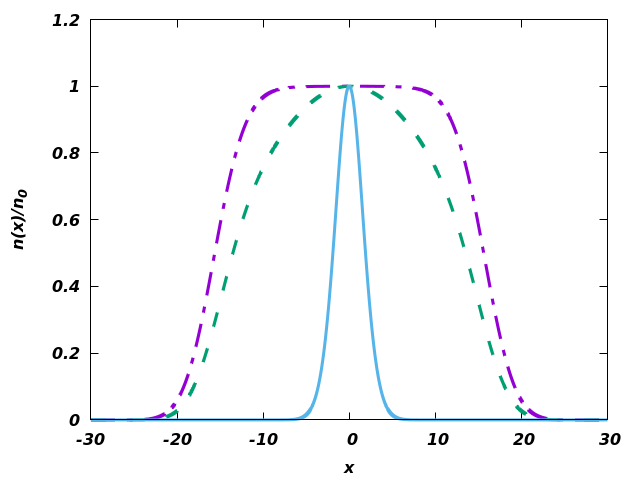}
\caption{(Color online) The density profile of the numerical solution for different $K$ value is depicted here. The purple dashed-dotted line described the density for $K=0.00001$, the green dashed line describes $K=0.001$ and the blue solid line is prepared for $K=0.1$. Here, $\alpha=g_2/|g_1|=0.1$ and $g_1$ is set at $1$.}\label{profile}
\end{figure}

The satisfactory comparison of our numerical result with the analytical solution motivates us to solve Eq.(\ref{bgp}) with $K\neq0$. We employ the imaginary time propagation method over small time steps ($5\times 10^{-6}$) for total $20000$ steps and solve the inhomogeneous CQNLSE for several trap frequencies and different BMF interaction strengths. We introduce the interaction strength in terms of $\alpha=g_2/|g_1|$ and we perform numerical calculations for three arbitrary values of $\alpha$, such that $\alpha=0.1, 0.5$, and $0.9$. Here, it must be noted that in Ref.\cite{debnath1}, the analytical solution was obtained for $\alpha=1/3$. In Fig.~\ref{profile} we depict the density distribution for $K=0.00001$ (purple dashed-dotted line), $K=0.001$ (green dashed line) and $K=0.1$ (blue solid line) respectively. We can clearly see that the density profile indicate towards a transition from a droplet to a soliton as the trap frequency is increased. At very low trap frequency, that is $K=0.00001$, the density plateau is distinctly visible indicating the droplet state. However, with gradual increase in trap frequency (say at $K=0.001$) we observe a bell like structure. A tighter confinement leads to the generation of bright soliton as described by the solid blue line. The figure was obtained for $\alpha=0.1$ however, we note that the profile is not significantly altered for larger values of $\alpha$ till $\alpha=0.9$. It is interesting to note that similar transition has already been observed for dipolar BEC where the dipolar interaction strength plays the analogous role of the trapping potential \cite{rafal}.

To excavate more on droplet-soliton transition, the chemical potential can be regarded as an important physical parameter. Hence, we assume, $\phi(x,t)=\psi(x)e^{-i\mu t}$ and apply in Eq.(\ref{bgp}). This results,
\begin{eqnarray}
\mu\psi(x)=\left[-\frac{1}{2}\frac{d^2}{dx^2}+\frac{1}{2}Kx^2+g_1\psi^2(x)+g_2\psi^3(x)\right]\psi(x).\nonumber\\
\end{eqnarray}
Applying the normalization condition $\int_{-\infty}^{\infty}\psi^2(x)dx=1$ we obtain,
\begin{eqnarray}
\mu=\int_{-\infty}^{\infty}\left[\frac{1}{2}\left(\frac{d\psi}{dx}\right)^2+\psi^2(x)\left(\frac{1}{2}Kx^2+g_1\psi^2(x)+g_2\psi^3(x)\right)\right].\nonumber\\
\end{eqnarray} 

\begin{figure}
\includegraphics[scale=0.35]{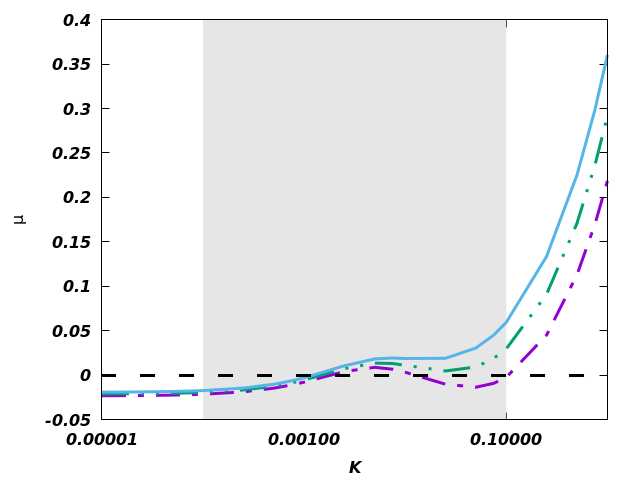}
\caption{(Color online) The variation of the chemical potential for different $\alpha$ is noted here. The solid blue line corresponds to $\alpha=0.9$, green dashed-double-dotted and purple dashed-dotted lines describe variation of the chemical potential with trap frequency for $\alpha=0.5$ and $0.1$ respectively. The black dashed line corresponds to $\mu=0$. The shaded area describes the region where we obtain a smooth transition from droplet to soliton.}\label{chem}
\end{figure}

We report the variation of the chemical potential with trap frequency for different beyond mean-field interaction strength in Fig~\ref{chem}. The purple dashed-dotted line, green dashed-double-dotted line and blue solid line depicts $\alpha=0.1, 0.5$ and $0.9$ respectively. For weaker trap frequency the system moves more in the droplet region which is signified by the negative value of the chemical potential. A very weak trap frequency also leads the system to a nearly homogeneous setup where we obtain the chemical potentials for different $\alpha$ is converging to $\sim -0.02$. Interestingly, similar result we have already reported for $\alpha=0.33$ \cite{debnath1}. As noted in Fig~\ref{profile}, the stronger confinement leads to the solitonic state with $\mu>0$ (right side of the grey shaded region of Fig~\ref{chem}). However, we observe a non-monotonic behaviour in the region $K=0.001$ and $0.1$ in the figure (the grey shaded region). A comparison with Fig~\ref{profile} encourages us to consider this frequency window as the transition region as we observe a bell like density profile for $K=0.001$. 

\begin{figure}
\includegraphics[scale=0.35]{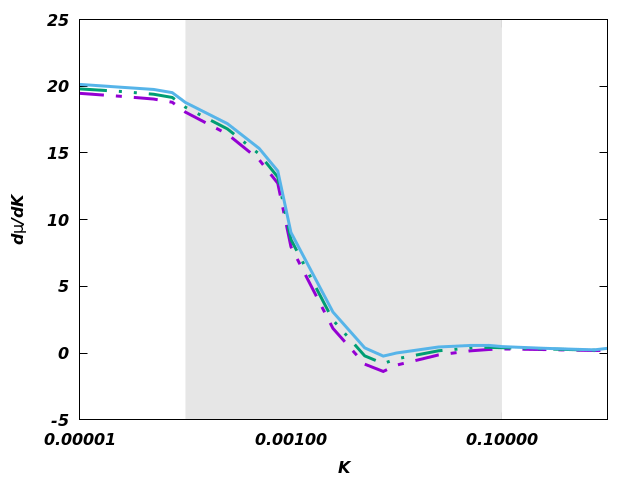}
\caption{(Color online) The first derivative of chemical potential w.r.t the trap frequency at different $\alpha$ is described. The color coding and the line types for three different $\alpha$ follows the same definition as of Fig~\ref{chem}. The shaded region describes the transition.}\label{dchem}
\end{figure}

To investigate the transition we then calculate the first derivative of chemical potential with respect to the trap frequency. The obtained result is depicted in Fig~\ref{dchem}. The main question here, is whether the transition is a phase transition (quantum) or a crossover. We know that, for quantum phase transition (QPT) we observe sudden variation in the ground state of the many body system when a controlling parameter ($\lambda$) of the Hamiltonian crosses a critical value \cite{khan1,khan2}. Here, we recognize $\lambda=K$, or the longitudinal trapping frequency. However, in Fig~\ref{chem} and Fig~\ref{dchem} we do not observe any abrupt change in the chemical potential. Thus, to our assessment the transition appears to be gradual and hence can be termed as a crossover. Based on these evidences we can also quantify the crossover region as $0.0001\lesssim K\lesssim 0.1$ (described in the grey shaded region in the figures). 

\begin{figure}
\includegraphics[scale=0.35]{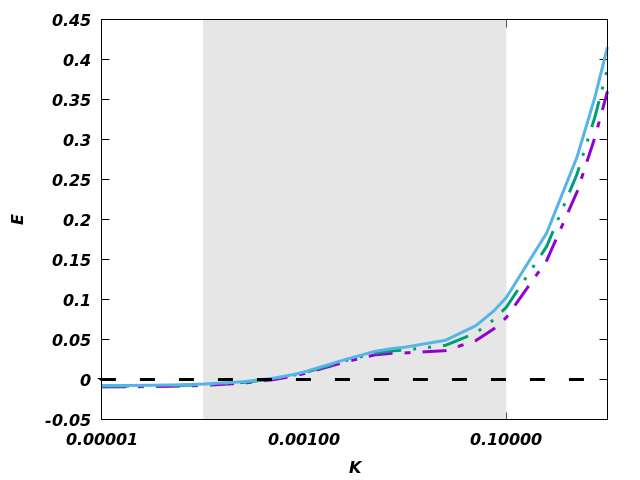}
\caption{(Color online) Change in energy with variation in the trap frequency is plotted here for different BMF interaction strengths. Again, the solid blue line, green dashed-double-dotted and purple dashed-dotted line denotes the energy variation as a function of trap frequency for $\alpha=0.9, 0.5$ and $0.1$ respectively. The black dashed line indicates $E=0$. The grey area corresponds to the region of transition from droplet to soliton.}\label{en}
\end{figure}

Further, we calculate the total energy which can be defined as,
\begin{eqnarray}\label{energy}
E&=&\int_{-\infty}^{\infty}\left[\frac{1}{2}\left(\frac{d\psi}{dx}\right)^2+\psi^2(x)\left(\frac{1}{2}Kx^2+\frac{1}{2}g_1\psi^2(x)\right.\right.\nonumber\\
&&\left.\left.+\frac{2}{5}g_2\psi^3(x)\right)\right].
\end{eqnarray}
The variation of energy corresponding to the trap frequency modulation is captured in Fig~\ref{en}. The figure reveals that the energy is negative for weak confinement suggesting the accumulation of droplet like bound pairs. The energy crosses the zero line in the vicinity of $K=0.001$ indicating the breakdown of droplet-like bound pairs. However, the energy flattening in the shaded area might be suggestive of an equilibrium of dropleton-soliton mixture. The bell shaped density profile can be noted as a signature of mixed type of states. Further tightening of the longitudinal frequency leads to the total destruction of the droplets.

\begin{figure}
\includegraphics[scale=0.35]{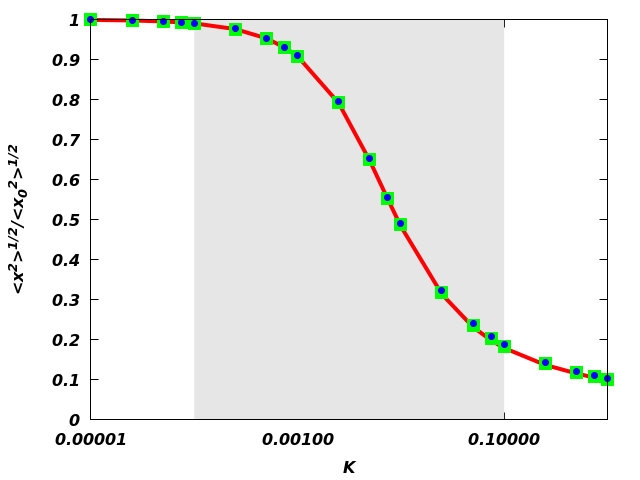}
\caption{(Color online) The figure describes the variation of the rms size of the droplets with the modulation of the trapping potential. The rms size is defined in units of $\sqrt{\langle x_0^2\rangle}$ where $\langle x_0^2\rangle$ is the rms size for $K=0$. The solid red line, the green solid squares and the blue solid circles denotes $\alpha=0.1$, $0.5$ and $0.9$ respectively. The shaded region describes the crossover area.}\label{rms}
\end{figure}

Next, we calculate the size variation in the droplet-soliton transition as depicted in Fig~\ref{rms} by calculating the root-mean-square (rms) ($\sqrt{\langle x^2\rangle}$) size of the condensate. The rms size is depicted in units of the rms size ($\sqrt{\langle x^2_0\rangle}$) for $K=0$. The figure clearly reflects on the fact that the variation of the BMF interaction does not have much contribution on the droplet size of the condensate as we report the variation for $\alpha=0.1$ (solid red line), $0.5$ (solid green squares) and $0.9$ (solid blue circles) respectively. We can also conclude that the weak confinement allows the larger pair size (with asymptotic limit extending to $1$) suggesting the droplet formation and gradually the pair size smoothly falls off as we increase the longitudinal frequency. We also observe that the pair size drops much faster in the shaded region, which we have defined as the crossover region. The rms curve flattens thereafter and asymptotically reaches $1/10$ of the droplet size for a tightly confined one dimensional geometry. The effect of the variation in $\alpha$ is apparently negligible for the normalized pair size, as noted in the figure.    
 
\begin{figure}
\includegraphics[scale=0.14]{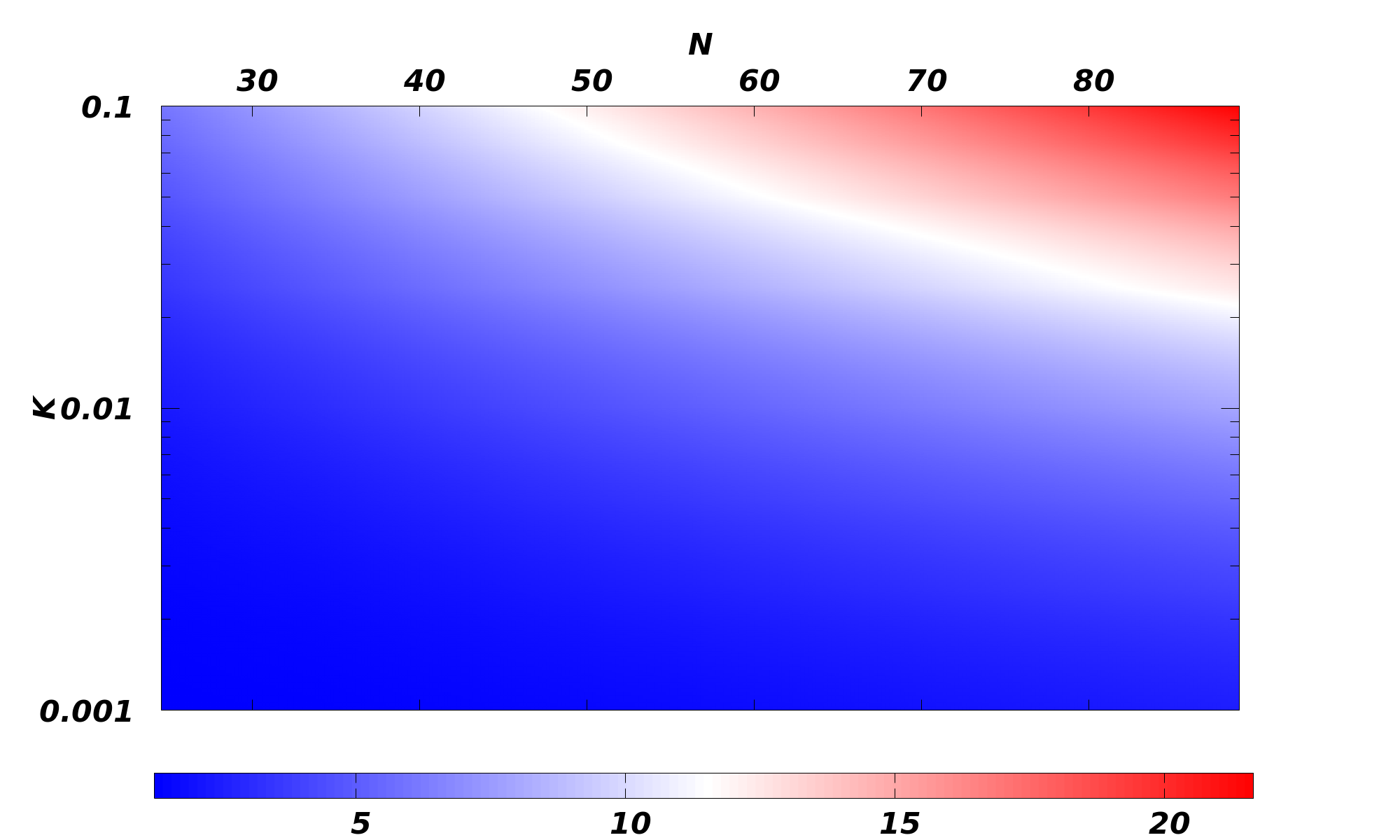}
\caption{(Color online) The figure describes the variation of the peak density as a function of particle number and $K$ for $\alpha=0.1$. A progressive phase separation can be noted.}\label{phase}
\end{figure}

\begin{figure}
\includegraphics[scale=0.35]{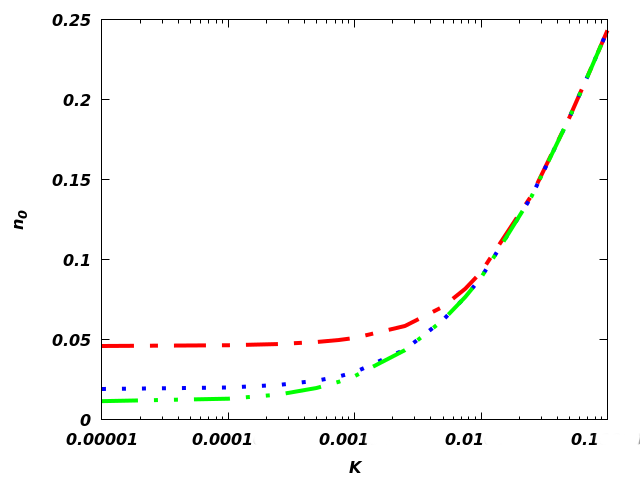}
\caption{(Color online) The figure describes the variation of the peak density ($n_0$) while the trapping potential is changed. The red dashed-dotted line describes $N=25$ while variation of $n_0$ for $N=50$ is depicted via blued dotted line. The green dashed-double-dotted line represents $N=80$. The plot is prepared for $\alpha=0.1$.}\label{n0}
\end{figure}

It is also instructive now to investigate the role of particle number of the condensate. It must be noted that peak density is a good indicator in understanding the droplet and the solitonic states as it does not change for droplets, whereas noticeable change can be viewed for a soliton. We draw a phase diagram for the peak density as a function of the longitudinal harmonic confinement and the particle number as depicted in Fig.~\ref{phase}. The extended blue region suggests of nearly constant peak density, thereby pointing towards the existence of droplet-like states. A relatively tighter confinement in the realm of Q1D geometry, we observe progressive changes in the peak density which suggests of an emergence of localized soliton-like states.
For further clarity we report the peak density $n_0$ for different trap potential in Fig~\ref{n0}. It can be clearly seen in the figure, that at weak trapping potential the density remains constant, thereby indicating the incompressible liquid-like state while progressive increase in the potential leads to the variation of the peak density suggesting higher compressibility which can be noted as a signature of gaseous soliton-like phase. The nature remains unchanged for different particle number as well as various values of $\alpha$ values.
\section{Conclusion}\label{con}
The recent experimental and theoretical developments related to the newly observed liquid like state in ultra-cold atomic BEC motivates us to study the role of harmonic confinement in the investigation of droplets.
Already we have shown the existence of analytical solution pertaining to a droplet-like state in a quasi one dimensional quasi homogeneous system \cite{debnath1}. Here, we start from where we left in the previous analytical description and 
explore the droplet state numerically in inhomogeneous quasi one dimensional system. We find that the longitudinal trap frequency can be an appropriate controlling parameter in experiments to access a crossover from a droplet to a solitonic mode. We observe, that at very weak trap frequencies, 
the droplet states do exists just like as they do in the homogeneous case. However, with systematic increase in the trapping potential we move from flat top droplet solution to bell shaped solution and later to the emergence of solitons. 
In this regard, we especially emphasize on the numerical calculation of the chemical potential and energy. For very weak trapping potential, the chemical potential converges to the analytical value reported earlier \cite{debnath1}.
At moderate values of the trapping frequency, the behavior of $\mu$ is nonlinear (the shaded regions in figure) and we observe a plateau in energy. Here, the droplet and soliton appear to be coexisting, corresponding to a bell type structure in the density.
At relatively tighter confinement, the bound droplets completely disappear and the solitonic mode prevails. We have repeated our investigation for different BMF interactions strengths ($\alpha=0.1$, $0.5$ and $0.9$) while noting a similar trend in the transition. To characterize the transition, we calculate the derivative of chemical potential however, we do not observe any significant change in the derivative as a function of the change of the Hamiltonian parameter ($K$ here). We also observe a smooth transition of the pair size across a weak to a strong confinement domain. This allows us to conclude that the droplet to soliton transition denotes  a crossover phenomena. 

As we report a confinement driven transition from droplet to soliton, we expect to open up new avenues in understanding the droplet state. We believe a study of fidelity of the numerically obtained solutions can shed more light in understanding the transition \cite{khan1}.
We also expect that our prediction can soon be verified in experiments.

\section*{Acknowledgement} AK acknowledge insightful discussion with P. Sadhukhan and P. Das. Authors also thank R. O\l dziejewski for valuable comments. AK also thanks Department of Science and Technology (DST), India
for the support provided through the project number CRG/2019/000108.

\bibliographystyle{apsrev4-1}
\bibliography{ms_v3}

\end{document}